# Black GeSn on Silicon for Enhanced Short-Wave Infrared Detection and Imaging

*Po-Rei Huang, Yue-Tong Jheng, and, Guo-En Chang\**

Po-Rei Huang, Yue-Tong Jheng, Guo-En Chang

Department of Mechanical Engineering, and Advanced Institute of Manufacturing with High-Tech Innovations (AIM-HI), National Chung Cheng University, Chiayi 62102, Taiwan
E-mail: imegec@ccu.edu.tw

**Abstract**

Sensitive and cost-effective Group IV short-wave infrared (SWIR) photodetectors (PDs), compatible with complementary metal–oxide semiconductor (CMOS) processes, are crucial for various emerging applications. Here, we developed a black GeSn thin-film PD on silicon, optimised for efficient SWIR photodetection and imaging. Incorporating Sn into the Ge layer effectively extended the photodetection range to 1960 nm. The blackening of the GeSn surface resulted in a substantial reduction of reflection loss across a broad spectral range of 1200–2200 nm. Furthermore, the responsivity experienced a remarkable increase of 145% through reduced reflection loss and carrier multiplication by the carrier ionization. The black GeSn surface significantly boosts the detectivity, enhances the wide-angle SWIR photodetection, and improves the quality of the resultant images. This demonstration heralds a new era for CMOS-compatible, cost-effective, high-performance black GeSn photodetectors and imagers for a wide range of applications in the underexplored SWIR region.

# 1. Introduction

Efficient and cost-effective infrared (IR) photodetectors (PDs), capable of operating in the short-wave infrared (SWIR, λ = 1400–2500 nm) region, have garnered considerable interest owing to their potential in diverse applications. These applications include fibre-optic telecommunications, integrated infrared cameras, automobile LiDAR sensing, spectroscopy, and night vision, to mention a few [1-3]. Currently, the predominant IR PDs in the market are primarily composed of III–V (e.g., InGaAs and InSb) and II–VI (e.g., HgCdTe) narrow-bandgap semiconductors [4]. Although these materials enable high-performance IR photodetection and imaging, they exhibit limited compatibility with standard complementary metal–oxide–semiconductor (CMOS) processes. This limitation hinders the monolithic integration of PDs with silicon-based readout circuits on a singular chip. Consequently, PDs and readout circuits are typically produced separately and subsequently connected via specialised flip-chip bonding techniques [5]. This approach incurs higher costs, increases fabrication complexity, and reduces yield rates. In response, research has pivoted towards developing CMOS-compatible SWIR PDs using group-IV semiconductors, facilitating their monolithic integration with circuits on a single chip, thus satisfying various application requirements. Significant advancements have been achieved in high-performance CMOS-compatible Si and Ge PDs and imagers [6]. However, owing to their relatively large bandgap energies, the cutoff wavelengths for Si and Ge PDs are confined to approximately 1100 nm and 1500 nm, respectively, posing challenges in extending their operational range to the SWIR region.

The recent breakthrough in developing group-IV GeSn alloys heralds a new era for low-cost, high-performance SWIR PDs. The strategic alloying of Ge with Sn effectively lowers the direct-bandgap energy, thus considerably extending the absorption edge from ~1500 nm toward longer wavelengths, enabling efficient photodetection in the infrared spectrum [7]. Despite the inherent limitation of only ~0.5% solid solubility of Sn in Ge [8], advancements in low-temperature growth techniques have enabled the growth of GeSn layers on Si or Si-on-insulator substrates with Sn concentrations up to 36% [9-12]. This significant progress has encouraged the development of diverse GeSn PDs, extending the photodetection range to as far as 4700 nm [13-23]. More importantly, recent demonstrations of GeSn image sensors [24-27] suggest immense potential for high-performance, cost-effective SWIR photodetection and imaging. However, enhancing the performance of GeSn PDs and imagers for practical applications remains a considerable challenge. A critical factor influencing the performance of GeSn PDs is their light collection efficiency. When light impinges on PDs from the far field, typically ambient air, a substantial reflection loss occurs owing to the refractive index (RI) mismatch between air and the semiconductor material. Fresnel equation dictates that if light is normally incident on the PD from the ambient air, the reflectivity (R) can be calculated using $R = (n − 1)^2/(n + 1)^2$, where n denotes the RI of the PDs. Consequently, approximately 30–40% of photons are reflected at the surface of the PDs [28], where the situation exacerbates at longer wavelengths due to the higher RI of narrow-gap semiconductors at longer wavelengths [29], resulting in increased reflection loss. Therefore, minimizing unwanted reflection is crucial for increasing light-collection efficiency, and thus, the performance of PDs.

Surface engineering of semiconductor PDs is considered a powerful strategy to substantially reduce reflection loss and improve PD performance [29-35]. Employing various nanostructures such as nanoholes [29], nanoneedles [32], and pyramids [33-35] on PD surfaces, through either bottom-up or top-down methods, can drastically reduce reflection loss. By creating nanostructures with dimensions smaller than the wavelength of incident light, exceptional light management can be achieved with significantly reduced reflection and enhanced absorption across a broad wavelength spectrum. Consequently, a greater number of incident photons are captured in the active layer and transformed into photocurrents, considerably amplifying the optical response of the PDs. Furthermore, this approach significantly reduces the manufacturing costs by eliminating the requirement for conventional, intricate, and time-intensive multilayer antireflection coatings. Prior researches have demonstrated several nanostructured surfaces, also referred to as black surfaces, using group-IV semiconductors. For instance, black Si surfaces fabricated from bulk Si substrates achieve reflection losses of less than 1% across a broad spectral range of 300–1000 nm; however, these losses notably increase at longer wavelengths [30]. Conversely, black Ge surfaces produced using bulk Ge substrates have attained low reflection (<2%) in the spectral range of 1000–2000 nm [36]. Critically, the nanostructures on these black surfaces provide opportunities for amplifying the photo-generated electron–hole pairs (EHPs), and thus photocurrents [31,36]. Therefore, when the black surfaces are applied to IR PDs, with their black surfaces and thick optical absorbers, they can achieve ultrahigh responsivity for sensitive photodetection in the near-infrared region [31,36]. Nevertheless, black Si and Ge PDs encounter inherent limitations regarding their detection range in the SWIR spectrum, higher material costs, and challenges in integration with other photonic and electronic devices. Yet, very little research has employed nanostructures to enhance the optical responses of GeSn PDs.

In this study, we demonstrate a thin-film-based black GeSn metal-semiconductor-metal (MSM) SWIR PD, monolithically integrated on silicon substrates, designed for efficient SWIR photodetection and imaging. The GeSn thin films were grown on silicon substrates through low-temperature growth techniques, which effectively decreased the bandgap energies, thereby extending the photodetection cutoff wavelength and enhancing the optical response in the SWIR region. The nanostructures were introduced using simple and cost-effective Cl2-based reactive ion etching (ICP-RIE) techniques to "blacken" the GeSn surface. This process significantly diminishes the reflection loss in the broad spectral range of 1200–2200 nm, consequently enhancing the optical responsivity. The responsivity is significantly improved in this spectral range of 1200–2200 nm, not only because of the reduced reflection loss but also owing to the current multiplication facilitated by the GeSn nanostructures. More importantly, the black GeSn PDs yielded enhanced wide-angle photodetection and SWIR images of superior quality. We posit that our black GeSn PDs will create new avenues for the development of high-performance, CMOS-compatible, and cost-effective SWIR PDs and imagers.

## 2. Results and Discussion
### 2.1 Fabrication of black GeSn photodetectors on silicon

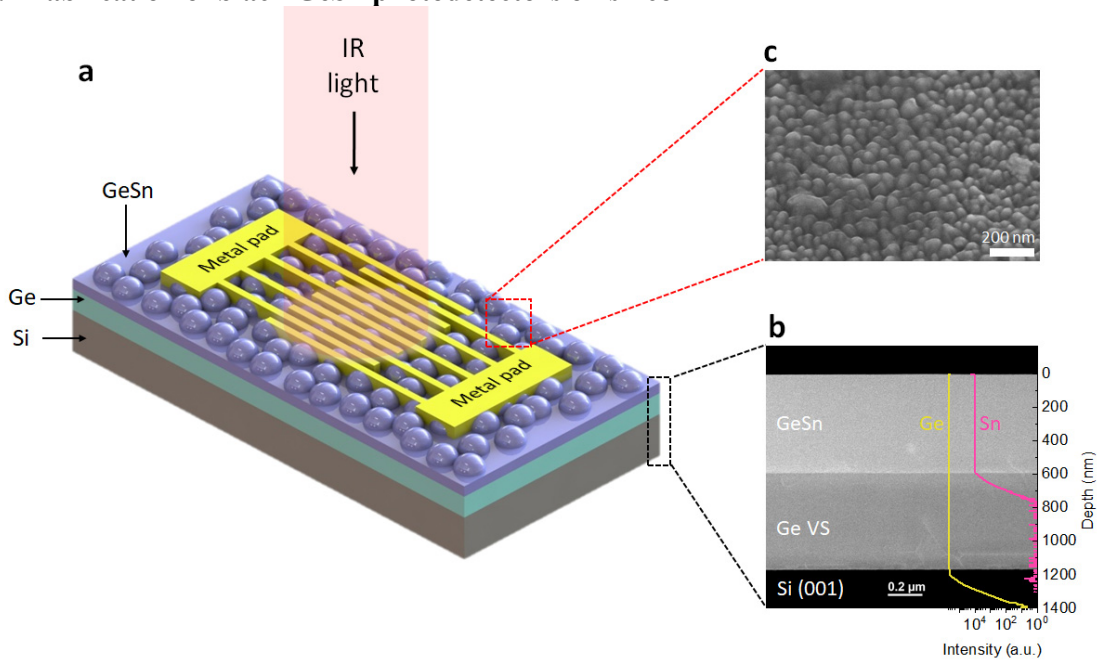

**Figure 1. Fabrication of black GeSn photodetectors on silicon.** (a) Schematic diagram of our black GeSn photodetectors on Si (not to scale). (b) Cross-sectional transmission electron microscopy (XTEM) image of the grown GeSn sample with an atomic element distribution of Ge and Sn atoms. (c) Scanning electron microscopy (SEM) image of the black GeSn surface.

The fabrication process of the black GeSn PDs involved material growth and device fabrication (Supplementary Information S1). A schematic diagram of the black GeSn PDs is illustrated in Fig. 1(a). The material for this study was grown on a silicon (001) substrate utilizing reduced-pressure chemical vapor deposition (RPCVD) along with low-temperature growth techniques, consisting a 620 nm Ge buffer layer and a 620 nm GeSn layer. Figure 1(b) portrays a cross-sectional transmission electron microscopy (XTEM) image of the grown sample, alongside the atomic distributions of Ge and Sn atoms, ascertained through secondary ion mass spectroscopy (SIMS) measurements. The XTEM image disclosed that misfit dislocations were localised at the interface between the Ge buffer and the Si substrate due to their lattice mismatch. The interface between the Ge buffer and the GeSn active layer was characterised by its clarity, sharpness, and flatness. SIMS atomic distribution indicated a uniform distribution of Sn atoms within the GeSn layers, devoid of significant Sn clusters or segregation. The Sn content and strain in the GeSn layer were quantified as 4.3% and –0.14%, respectively, employing X-ray diffraction techniques (Supplementary Information S2.).

To fabricate the black GeSn surfaces, the grown GeSn sample underwent ICP-RIE process in a $Cl_2$ environment, optimizely conducted for 120 s to create nanostructures on the surface of the grown GeSn sample. The surface morphology of the GeSn nanostructures was examined using scanning electron microscopy (SEM) and atomic force microscopy (AFM). Figure 1(c) show an SEM image of the black GeSn surface, showing that the resulting random GeSn nanostructures exhibited sizes ranging from several tens to several hundred nanometres. Note that the dimensions and aspect ratio of the GeSn nanostructures are closely associated with the ICP-RIE etching parameters, including chamber pressure, etching duration, RF power, and $Cl_2$ flow rate, as well as the thickness of the GeSn layer. Excessive etching could result in the complete removal of the GeSn film, while insufficient etching time could yield inadequate height and aspect ratio of the GeSn nanostructures. AFM results revealed the average height and width of the GeSn spikes to be 44 nm and 130 nm, respectively, with the residual GeSn film of thickness of 297 nm (Supplementary Information S3). The samples were then fabricated into MSM PDs with interdigitated metal pads (length: 700 μm; width and pitch: 60 μm) for optoelectronic evaluation. For comparative purposes, a control GeSn PD device was fabricated using the as-grown GeSn sample, maintaining the same device design (labelled as planar GeSn PD).

## 2.2 Reflectivity of black GeSn surfaces

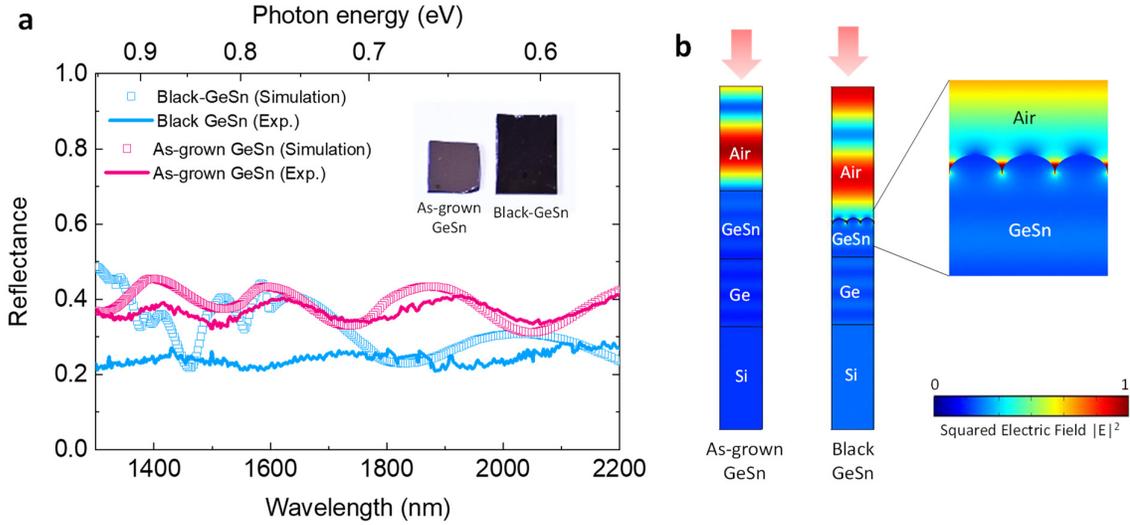

**Figure 2.** Experimental and theoretical verification of reflectivity for black GeSn surfaces on silicon. (a) Measured and simulated reflectivity spectra of as-grown GeSn sample and black GeSn. Inset: optical images of as-grown and black-GeSn samples. (b) Simulated field distribution of as-grown GeSn sample and black GeSn at $\lambda$=1580 nm.

To assess the effectiveness of GeSn nanostructures in reducing reflectivity, the surface reflectivity of both as-grown GeSn sample and black GeSn surfaces was examined. This involved experimental reflection measurements and comparison with theoretical simulations using the finite element method (FEM) simulations (Methods). (In the FEM simulations, the GeSn nanostructures were presumed to be uniform and periodic.) Figure 2(a) displays the measured and simulated reflectance spectra for the as-grown and black GeSn samples. The as-grown GeSn sample exhibited a high average reflectance of ~0.365 owing to the significant RI mismatch between the air and semiconductor, with subtly varying ripple structures attributed to interference between the layers, and a free spectral range of ~300 nm. The reflectivity also increased at longer wavelengths owing to the increased RI. In contrast, the black GeSn sample demonstrated a significantly reduced reflectance of ~0.24, with a larger free spectral range, attributable to geometric light trapping by the GeSn nanostructures. The inset in Fig. 2(a) compares the optical images of the as-grown and black GeSn samples. The black GeSn sample appeared darker than its as-grown counterpart, signifying reduced light reflection at the black GeSn surface. Notably, the simulated reflectance spectrum of the as-grown GeSn sample closely aligns with the measured spectrum. However, the simulated reflectance for black GeSn is higher than the measured spectrum, because actual black surfaces with nonuniform and random structures exhibit superior antireflection effects [36,37].

To delve deeper into the impact of GeSn nanostructures on antireflection, we analysed the electromagnetic field distribution and antireflection behaviour of both as-grown and black GeSn samples. Figure 2(b) presents the simulated squared electric field distribution for as-grown and black GeSn with uniform GeSn nanostructures at λ = 1580 nm. For the as-grown GeSn sample, standing-wave patterns were discernible in the GeSn and Ge buffer layers because of the interference in the films resulting from the reflection of light at the Ge/Si interface. Moreover, a pronounced intensity was observed above the GeSn surface because of significant light reflection. In contrast, the black GeSn sample showed a reduction in intensity above the GeSn nanostructure, reflecting diminished reflection. The intensity focused between the GeSn nanostructures, suggesting multiple reflection effects that intensified the light. The amount of light within the black GeSn structure was increased, indicating enhanced light penetration into the active GeSn layer. Furthermore, the intensity of the transmitted light in the Si layer decreased, signifying greater light absorption by the GeSn layer.

**2.3 Electrical and optical characterization of black GeSn PDs**

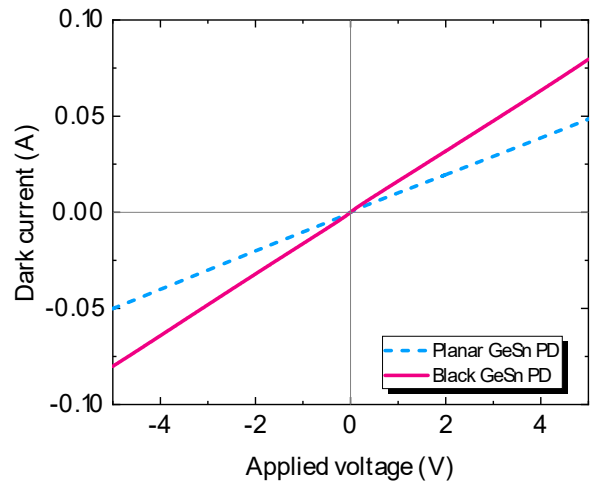

**Figure 3.** Dark current–voltage characteristics for planar and black GeSn PDs.

To investigate the electronic and optical characteristics of the black GeSn PDs and their reference planar counterparts, experiments were conducted on devices both with and without illumination. Figure 3 presents the current–voltage ($I_{dark}$–$V$) characteristics of the reference planar GeSn PD and the black GeSn PD in dark conditions. For both devices, the dark current demonstrated a linear increase with escalating applied bias voltage, indicative of an ohmic contact between the GeSn and metal pads. Notably, the black GeSn PD exhibited a higher dark current compared to the reference planar GeSn PD, likely attributed to its larger surface area that enhances carrier generation, recombination, and tunnelling [30].

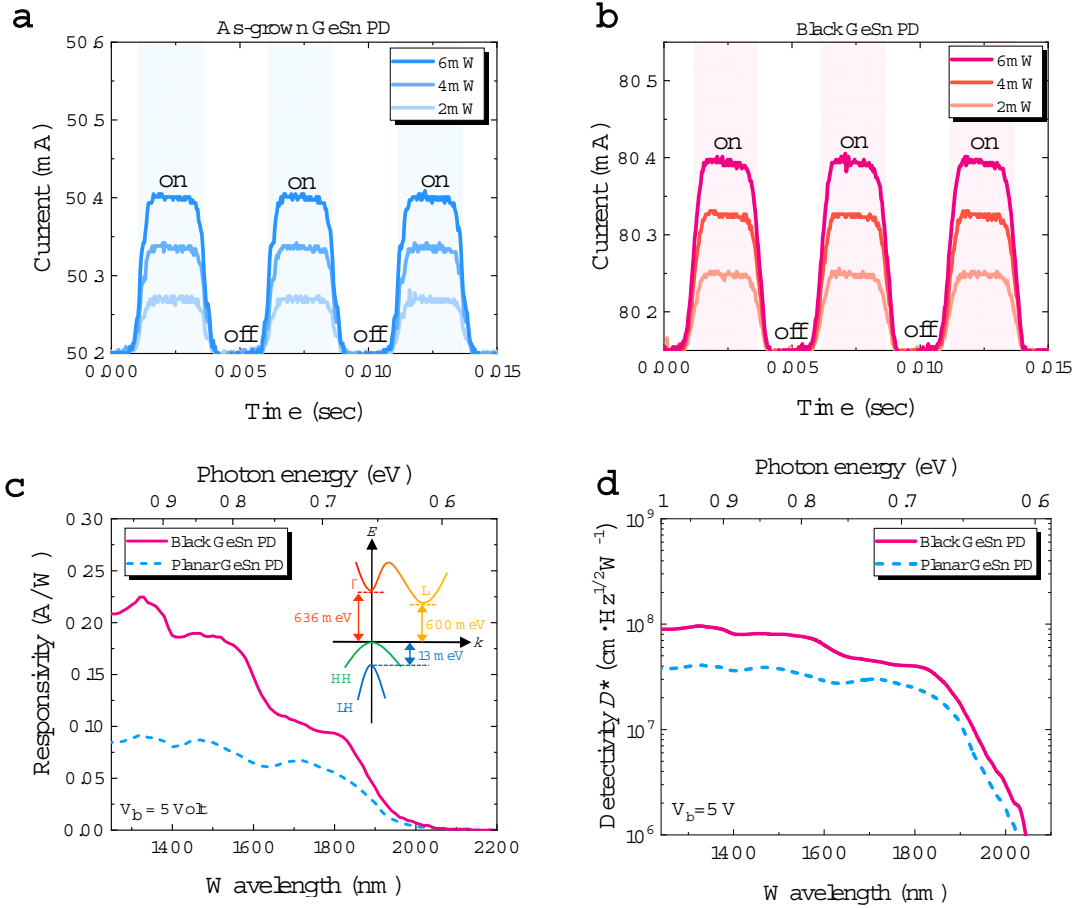

**Figure 4. Optical characterization of black GeSn photodetectors.** Time-responses for (a) reference planar and (b) black GeSn photodetectors biased at 5 V with illumination using a 1900 nm light source with various optical powers. (c) Room-temperature responsivity spectra biased at 5 V. The inset shows the calculated band structure of the GeSn active layer. (d) Room-temperature detectivity spectra biased at 5 V.

Figure 4(a) and 4(b) depict the temporal responses of the reference planar and black GeSn PDs, respectively, when exposed to a 1900 nm broadband light source (ASE 1900, Thorlabs) chopped at 200 Hz. Both devices displayed discernible switching behaviour. The photocurrent in each device increased with the intensity of the incident optical power, as higher optical power generates more electron-hole pairs, resulting in larger photocurrents. These observed photocurrents further substantiate the capability of GeSn PDs for SWIR photodetection.

Despite the relatively minimal thickness of the GeSn layers, the black GeSn PDs exhibited significantly higher responsivity compared to the reference planar GeSn PD. Figure 4(c) illustrates the measured optical responsivity spectra of the black GeSn PD and the reference planar GeSn PD at a 5 V bias voltage. The responsivity decreases with the increasing wavelength and became negligible at λ = 1960 nm in both devices, aligning with the direct-gap absorption edge (the direct-bandgap energy of the material). The photodetection range was notably extended beyond that of pure Ge (~1500 nm) owing to the addition of Sn. To verify the direct-gap energy, we computed the band structure of the GeSn active layer using deformation potential theory, accounting for strain effects [38]. A schematic of this calculated band structure is depicted in the inset of Fig. 4(c). The compressive strain splits the degenerate valence band, elevating the heavy-hole (HH) band above the light-hole (LH) band. The incorporation of Sn into the active layer lowers the Γ-conduction band, thereby defining the lowest direct-bandgap by the optical transition energy from the HH band to the Γ-conduction band (HH → Γc transition). We determined a transition energy of 636 meV for the HH → Γc transition, closely aligning with the experimental result of 632 meV, thereby confirming the extension of direct-gap absorption by Sn alloying. Intriguingly, when comparing the responsivity of the devices, the responsivity of the black GeSn PD was approximately 145% higher than that of the reference planar GeSn PD, despite its smaller thickness (320 nm for the black GeSn PD versus 620 nm for the reference planar GeSn PD). This suggests a current gain in the black GeSn PD, likely due to impact ionisation occurring within the GeSn nanostructures, a phenomenon previously observed in other group-IV-based black surfaces, including Si [31] and Ge [36].

The overall performance of the PDs can be quantitatively evaluated by the specific detectivity ($D^*$), which can be derived from the dark current and responsivity. The specific detectivity spectra of both the reference planar and black GeSn PDs are presented in Fig. 4(d). The as-grown GeSn PD exhibited a $D^* \approx 3.5 \times 10^7 \, \text{cm} \cdot \text{Hz}^{1/2} \text{W}^{-1}$ in the spectral range of 1250–1800 nm, with a decline commencing at ~1900 nm. However, the black GeSn PD, despite its higher dark current, demonstrated an enhanced $D^* \approx 7.5 \times 10^7 \, \text{cm} \cdot \text{Hz}^{1/2} \text{W}^{-1}$, corresponding to a notable improvement of 114%.

## 2.4 Angle-dependence responsivity

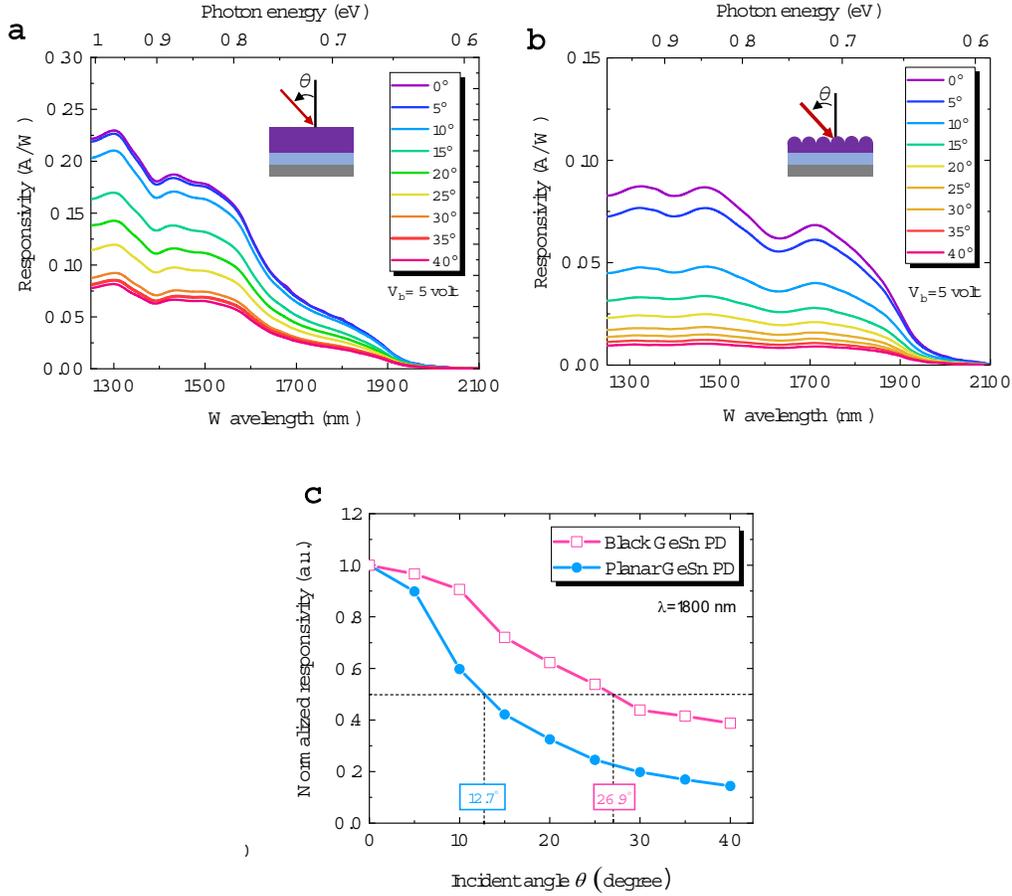

**Figure 5.** Optical responsivity spectra for (a) reference planar and black GeSn photodetectors biased at 5 V with various angles of incidence. (c) Normalised responsivity at λ=1800 nm as a function of incident angle.

Wide-angle photodetection is crucial for both photodetection and imaging applications. To assess the wide-angle photodetection capabilities, responsivity experiments were performed at varying incident light angles ($\theta$) ranging from 0° to 40° for both the reference planar GeSn and black GeSn PDs, biased at 5 V. The outcomes of these experiments are depicted in Figure 5(a) and 5(b). In each scenario, the responsivity deteriorated with an increase in the incident light angle. Figure 5(c) illustrates the normalised responsivity at λ = 1800 nm for both reference planar and black GeSn PDs as a function of the incident light angle. The responsivity exhibited a decrease with an escalating incident light angle, aligning with the cosine law [39,40]. The responsivity of the reference as-grown GeSn and black GeSn PDs decreased to 50% at an incident angle of $\theta$ = 12.7°. Conversely, the responsivity of the black GeSn PD declined more gradually, reaching 50% at an incident angle of $\theta$ = 26.9°. This finding indicates that the GeSn nanostructures are effective in reducing reflection loss and capturing more incident photons into the GeSn active layer compared to the reference planar GeSn PD. These findings suggest that the black GeSn PD is capable of achieving superior wide-angle photodetection performance relative to planar GeSn PDs under inclined conditions.

## 2.5 Short-wave infrared imaging

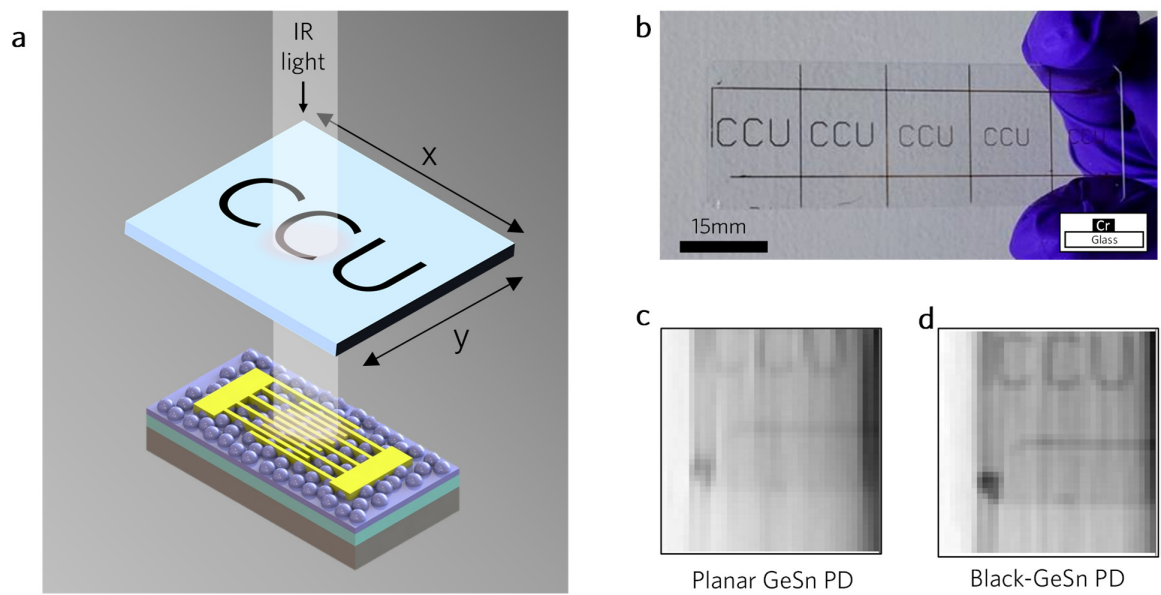

**Figure 6. SWIR imaging results.** (a) Schematics of single-pixel infrared imaging experiments (not to scale). (b) Optical image of the glass mask as the target. SWIR images captured by (c) reference planar GeSn PD and (d) black-GeSn PD under uncooled conditions.

Single-pixel imaging experiments were conducted to explore the potential of the fabricated black GeSn PD for SWIR imaging applications. The experimental setup is illustrated in Fig. 6(a) and further detailed in Supplementary Information S4. A glass substrate with Cr patterns was utilised as the mask (Fig. 6(b)). Figure 6(c) presents a comparative analysis of the SWIR images captured at room temperature using a broadband SWIR source. Both the as-grown GeSn and black GeSn PDs successfully acquired infrared images. However, the image quality from the reference planar GeSn PD appeared blurry, whereas the image from the black GeSn PD was markedly clearer and sharper, attributable to enhanced detectivity. These outcomes underscore the advantages of black GeSn PDs as CMOS-compatible SWIR imagers.

## 3. Conclusion

In summary, this study presents a novel black GeSn thin-film photodetector on silicon, offering unique CMOS-compatibility and efficient SWIR photodetection and imaging. Incorporating 4.3% Sn into the GeSn active layer markedly extends the photodetection range within the SWIR spectral region due to the reduced bandgap energy. The formation of GeSn nanostructures, achieved through rapid and cost-effective dry etching techniques, effectively blackens the GeSn surface. This process substantially diminishes reflection loss across a wide spectral range of 1200–2000 nm, corroborating our theoretical predictions. Furthermore, these nanostructures facilitate carrier multiplication, significantly boosting both responsivity and detectivity. Consequently, black GeSn PDs exhibit enhanced wide-angle photodetection and superior image quality compared to traditional planar GeSn PDs. Future enhancements in photodetection and imaging performance are expected by increasing detectivity.

This improvement can be realised by optimizing GeSn nanostructures and increasing the thickness of the GeSn active layer. Additionally, minimizing dark currents through effective passivation and/or *p–i–n* diode technologies is critical. Increasing the Sn content within the GeSn active layer may also extend the photodetection range of the device into the mid-infrared spectrum. This study demonstrates a straightforward yet impactful methodology for developing efficient GeSn photodetectors and imagers on silicon, thereby forging a new path for CMOS-compatible SWIR photodetectors and imagers suitable for a broad array of applications.

## 4.. Experimental Section/Methods

*FEM Simulation*

3D FEM simulations were executed to evaluate the optical behaviour of both as-grown and black GeSn samples. Wavelength-dependent refractive indices and extinction coefficients for the materials were sourced from Refs. [7,40]. A plane wave functioned as the light source incident on the device, and the electromagnetic field distribution was then ontained. The simulations provided data on field distribution, transmittivity ($T$), and reflectivity ($R$). The absorbance ($A$) was calculated using the relation $A = 1 - R - T$.

*Optical responsivity*

The optical responsivity of the devices was ascertained using a tunable light source (TLS130B-250Q, Newport), which was optically chopped at 200 Hz, was reshaped through an adjustable pinhole, and then directed to be incident normally on the GeSn PDs. The GeSn PDs were connected in series with a load resistance of 50 Ω and biased using a sourcemeter (Keithley 2400). The resultant voltage drop across the load resistance was then read using a lock-amplifier (SR830, Stanford Research System) for analysis of the photocurrent spectrum. The incident optical power was measured using an optical power meter (S148C, Thorlabs). The responsivity spectra were then determined from the measured photocurrent and incident optical power.

**Acknowledgements**

This work was supported by the Young Scholar Fellowship Program of the National Science and Technology Council of Taiwan (NSTC) under Grant Nos. NSTC 112-2636-E-194-001 and NSTC 113-2636-E-194 -001, and the Taiwan Semiconductor Research Institute (TSRI) under Grant No. JDP112-Y1-028. The authors would like to thank Prof. Shang-Hua Yang at National Tsing Hua University, Taiwan, for assistance with imaging experiments.

Supporting Information for

**Black GeSn on Silicon for Enhanced Short-Wave Infrared Detection and Imaging**

*Po-Rei Huang, Yue-Tong Jheng, and, Guo-En Chang\**

Po-Rei Huang, Yue-Tong Jheng, Guo-En Chang

Department of Mechanical Engineering, and Advanced Institute of Manufacturing with High-Tech Innovations (AIM-HI), National Chung Cheng University, Chiayi 62102, Taiwan
E-mail: imegec@ccu.edu.tw

**Contents:**

**S1: Growth and fabrication of black GeSn photodetectors on Si**

**S2: Determination of Sn content and strain**

**S3: Morphology of GeSn nanostructures**

**S4: Experiment setup for single-pixel imaging**

## S1: Growth and fabrication of black GeSn photodetectors on Si

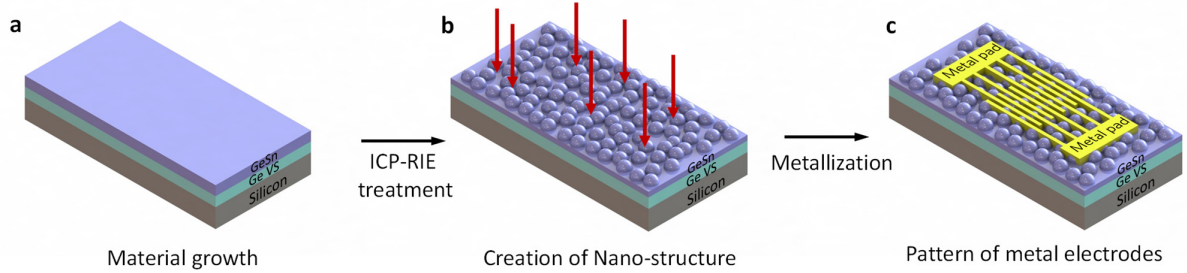

**Figure S1.** Fabrication flow of black GeSn PDs on a Si(001) substrate. (a) Growth of GeSn sample on a silicon substrate with a Ge buffer, (b) creation of GeSn nanostructures using dry etching techniques, and (c) deposition and patterning of metal electrodes.

The fabrication flow of the black GeSn PD is depicted in Fig. S1. Regarding material growth (Fig. S1(a)), the material utilised in this study was grown using reduced-pressure chemical vapor deposition (RPCVD) with $Ge_2H_6$ and $SnCl_4$ precursors. The epitaxial process started with the formation of a strain-relaxed Ge buffer layer through a two-step growth technique, incorporating an *in situ* annealing step at 800 °C to improve the material quality. This was followed by the growth of a 640 nm GeSn layer at 320 °C to inhibit Sn segregation. To fabricate the black GeSn surfaces growth (Fig. S2(a)), the grown GeSn sample underwent a cleaning process with methanol, acetone, and DI water. This was followed by ICP-RIE in a $Cl_2$ environment, conducted for 120 s to create nanostructures on the surface of the grown GeSn sample. The MSM GeSn PDs were fabricated utilizing lift-off techniques (Fig. S1(c)). Initially, patterns for interdigitated pads (length: 700 μm; width and pitch: 60 μm) were created using standard optical lithography. Subsequently, Au/Cu bilayers (200/20 nm) were deposited through an electron-beam (e-beam) evaporator. The formation of interdigitated metal pads was completed by immersing the samples in acetone, which served to eliminate any residual photoresist.

## S2. Determination of Sn content and strain

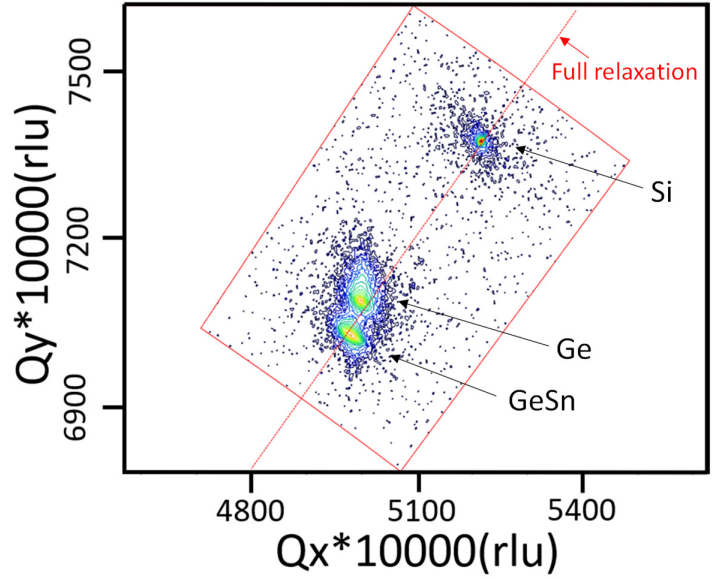

**Figure S2.** XRDRSM contour plot of the grown GeSn sample.

The structural characteristics of the developed GeSn sample were meticulously examined using X-ray diffraction reciprocal space mapping (XRDRSM) techniques. Figure S2 presents the XRDRSM of the 224 plane. Three prominent peaks were identifiable, corresponding respectively to the Si substrate, the Ge virtual substrate (VS), and the GeSn layer. Notably, the Ge peak appeared slightly above the strain-relaxation line, suggesting the presence of tensile strain within the Ge VS owing to the mismatch between the thermal expansion coefficients of Si and Ge during the annealing phase of epitaxial growth [S1]. Conversely, the GeSn peak was located below the strain-relaxation line, indicative of compressive strain. The in-plane ($a^{\parallel}$) and out-of-plane ($a^{\perp}$) lattice constants of these layers were deduced from the positions of these peaks. Subsequently, the bulk lattice constant ($a_0$) was derived using:

$$a_0 = \frac{a^{\perp} + 2a^{\parallel}\mu}{1+2\mu} \tag{S1}$$

where $\mu = 0.3738 + 0.1676x - 0.0296x^2$ for $Ge_{1-x}Sn_x$ [S2]. The composition-dependent lattice constants of the bulk GeSn were obtained as follows:

$$a_0(Ge_{1-x}Sn_x) = (1-x)a_0(Ge) + xa_0(Sn) \tag{S2}$$

where $a_0(\text{Ge}) = 5.6573$ Å and $a_0(\text{Sn}) = 6.4892$ Å are the bulk lattice constants of Ge and α-Sn, respectively [S3]. Comparing Eqs. (S1) and (S2) yielded an Sn content of the GeSn active layer $x = 4.3\%$. The in-plane biaxial strain in the GeSn layer can be obtained as

$$\varepsilon^{\parallel} = \frac{a^{\parallel} - a_0}{a_0} \tag{S3}$$

We obtained $\varepsilon^{\parallel} = -0.14\%$ a for the GeSn layer.

**S3: Morphology of GeSn nanostructures**

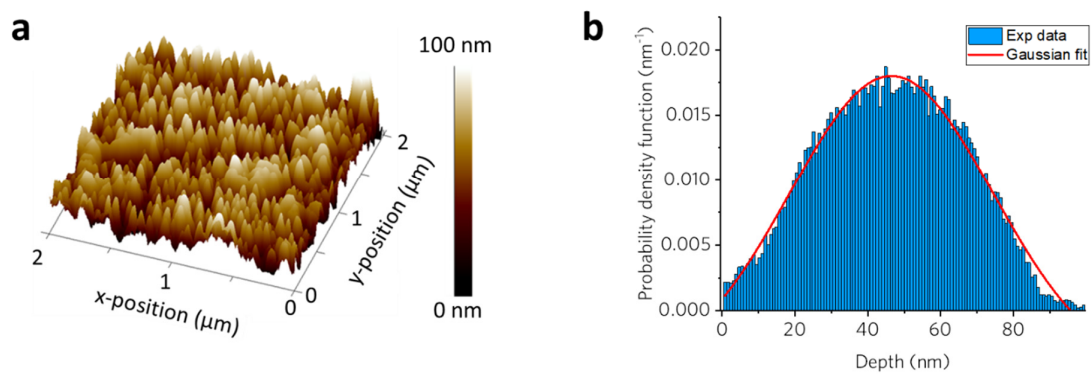

**Figure S3.** (a) AFM image of the grown GeSn sample. (b) Probability density function.

Figure S3(a) shows an atomic force microscopy (AFM) image of the black GeSn surface. Random GeSn nanostructures were clearly observed. Figure S3(b) illustrates the probability density function of the black GeSn surface, exhibiting a Gaussian-like distribution. Applying a Gaussian fit to the data, the average height of the GeSn nanostructures was determined to be 44 nm. The average width of 130 nm for these nanostructures are obtained from the AFM results.

**S4: Experiment setup for single-pixel imaging**

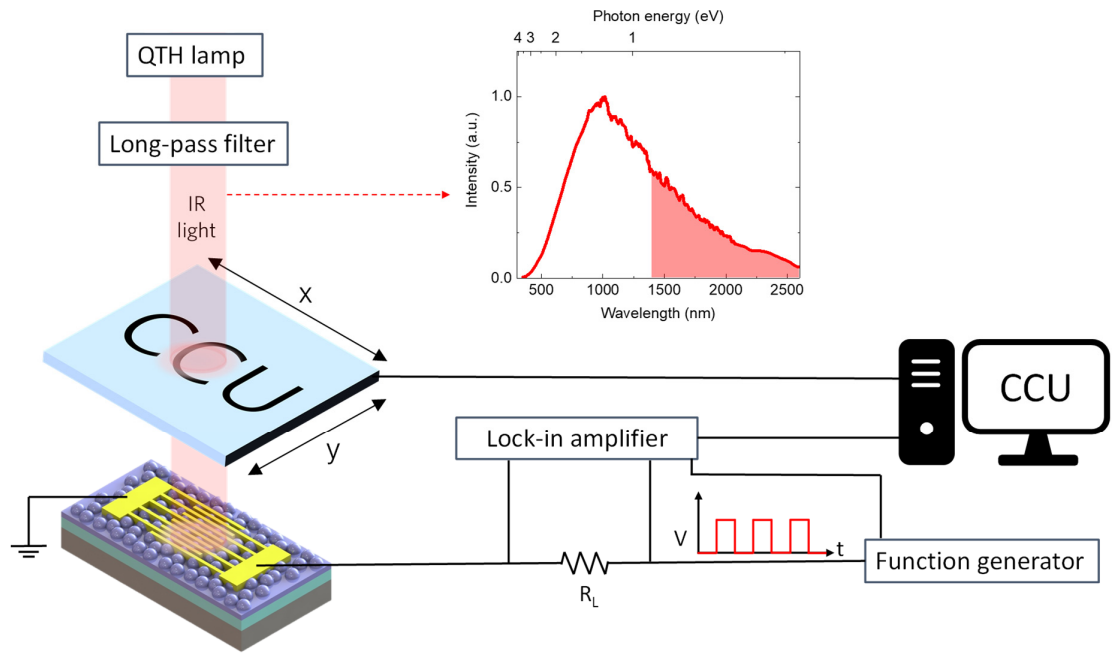

**Figure S4**. Setup of single-pixel infrared imaging experiments.

Figure S4 shows the setup of the SWIR imaging experiments. In the SWIR imaging experiments, we utilised a stabilised tungsten halogen light source (360–2600 nm) (Thorlabs, SLS201L) with a 1400 nm long-pass filter. The light passed through a mask positioned on a motorised XY stage before being incident normally on the GeSn PD, which was connected in series with a load resistance ($R_L$ = 50 Ω). A function generator supplied AC voltage with a square wave at a specified frequency (200 Hz) to power the GeSn PD. The voltage drop across the load resistance was quantified using a lock-in amplifier (Stanford Research Systems, SR 830). This setup enabled the analysis of the position-dependent photocurrent from which the infrared images were obtained.